\documentclass[twocolumn,showpacs,preprintnumbers,amsmath,amssymb]{revtex4}
\usepackage{graphicx}% Include figure files
\usepackage{dcolumn}% Align table columns on decimal point
\usepackage{bm}% bold math

\begin{document}

\preprint{APS/123-QED}

\title{Radio-frequency reflectometry on large gated 2-dimensional systems}% Force line breaks with \\

\author{L.J. Taskinen}
\author{R.P. Starrett}
\author{T.P. Martin}
\author{A.P. Micolich}
\author{A.R. Hamilton}
\author{M.Y. Simmons}
\affiliation{School of Physics, University of New South Wales, Sydney NSW 2052, Australia}
\author{D.A. Ritchie}
\author{M. Pepper}
\affiliation{Cavendish Laboratory, University of Cambridge, Cambridge CB3 0HE, United Kingdom}

\date{\today}% It is always \today, today,

\begin{abstract}
We have embedded an AlGaAs/GaAs based, gated 2D hole system (2DHS) into an impedance transformer $LC$ circuit, and show that by using radio-frequency reflectometry it is possible to perform sensitive, large bandwidth, electrical resistance measurements of 2D systems at mK temperatures. We construct a simple lumped element model where the gated 2DHS is described as a resistive transmission line. The model gives a qualitative understanding of the experimental results. As an example, we use our method to map out the Landau level evolution in a 2DHS as a function of magnetic field and gate voltage. 
%Valid PACS numbers may be entered using the \verb+\pacs{#1}+ command.
\end{abstract}

\maketitle

\section{\label{sec:intro}Introduction}
A large bandwidth in electrical measurements of mesoscopic and low dimensional devices enables interesting experiments. For example, shot noise measurements can yield important information on the transport properties of these systems \cite{blan00}. There is a growing interest in studying resistance relaxation and noise properties of 2D systems, for example the still controversial metal insulator transition \cite{krav04} and glassy dynamics \cite{jar06} could benefit from fast and low noise measurement techniques. 

The problem in cryogenic measurements is the need for long wiring connecting the sample at low temperature to the room temperature amplifiers. This together with the large resistance of typical samples leads to a large $RC$ time constant and limits the measurement bandwidth. One way to overcome this problem is to use a low temperature amplifier close to the sample \cite{vink07}, but such amplifiers have problems with large 1/$f$ noise and there are limits to how close to the sample it can be installed due to heat dissipation in the amplifier.

Another well known method to increase the bandwidth is to embed the sample into an impedance matching $LC$ circuit that terminates a transmission line connected to the room temperature measurement setup and use radio-frequency (rf) reflectometry to measure temporal changes in the resistance of the devices. Previously this technique has been used to increase the measurement bandwidth of single electron transistor \cite{schoe98} and quantum point contact \cite{qin06, cass07} electrometers, and superconductor-insulator-normal thermometers \cite{schm03}. Initially one would expect that it is not possible to use the rf reflectometry technique with large area gated 2D systems because the large capacitance results in a negligible sensitivity at high resistances. In this paper we show that it is possible to perform high sensitivity rf reflectometry measurements of large area 2D systems, and construct a simple model to explain how this works. The rf measurement can be calibrated by performing simultaneous four-terminal a.c. lock-in and rf reflectometry measurements.

\section{\label{sec:RFintro}Reflectometry technique and expectations for 2D devices}
\begin{figure}[h]
\includegraphics[width=0.95\linewidth]{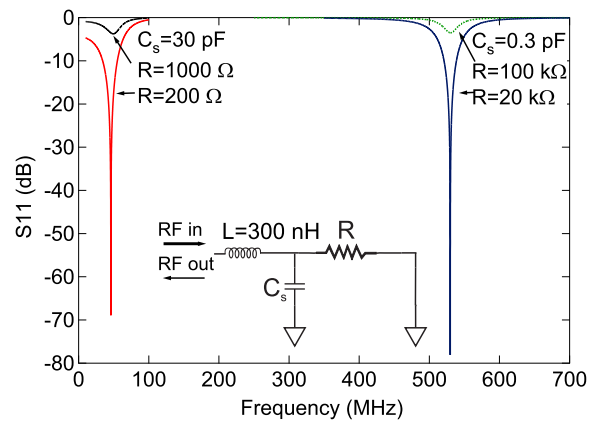}% Here is how to import EPS art
\caption{\label{fig:basicRLC} Simulated rf reflectance $S11$ for two different stray capacitances $C_s=30$ pF and $C_s=0.3$ pF. The inductance is $L=300$ nH in both cases. The inset shows the $RLC$ circuit used in the simulations.}
\end{figure}
The samples we want to study typically have a high resistance of the order of the resistance quantum $R_Q=25.8$ k$\Omega$ or higher. If one terminates a 50 $\Omega$ transmission line with such a sample, the large mismatch in the two impedances makes it impossible to perform high sensitivity and high bandwidth measurements. The large resistance $R$ of the sample can be transformed close to the impedance of the transmission line $Z_0$ (usually $Z_0=50$ $\Omega$) by using an impedance matching circuit, as shown in the inset of Figure \ref{fig:basicRLC}. The matching circuit consists of an inductor $L$, typically a surface mounted component placed close to the sample, and the capacitance $C_s$ which includes the stray capacitance from the inductor and the sample pads to ground. The inductance $L$ and capacitance $C_s$ form an $LC$ resonator with  resonance frequency $f_c=1/2\pi LC_s$. At the resonance frequency the impedance of the matching circuit is real and is given by $Z_t=L/RC_s$. Perfect matching occurs when $Z_t=L/R_mC_s=Z_0$, where $R_m$ denotes the matching resistance. Variations in the sample resistance change the transformed impedance $Z_{t}$ and hence the reflection coefficient $\Gamma=(Z_t-Z_0)/(Z_t+Z_0)$. Then, instead of measuring the resistance $R$ directly, one launches a rf carrier wave down the transmission line at a frequency close to $f_c$, where the sensitivity to the changes in $R$ is largest, and monitors the amplitude of the reflected carrier wave. Since the carrier frequency is typically a few hundred megahertz or higher, this method not only increases the bandwidth, but avoids the amplifier 1/$f$ noise.

The bandwidth of the circuit is $f_c/Q$ where $Q$ is the quality factor of the measurement setup. The quality factor for the matching circuit has two contributions, the unloaded quality factor which depends on the sample resistance and is defined as $Q_u=R/\sqrt{L/C}$, and the external quality factor $Q_e=\sqrt{L/C}/Z_0$. The total quality factor is given by $Q=\left(Q_u^{-1}+Q_e^{-1}\right)^{-1}$. At resistances larger than the matching resistance, $R>R_m$, the sample is overcoupled to the transmission line and the bandwidth is set by $Q_e$. At small sample resistances, $R<R_m$, the sample is said to be undercoupled to the transmission line and $Q\approx Q_u$. A more thorough introduction to the reflectometry technique can be found in Ref. \cite{ros04}.    

The reflectometry method works well with mesoscopic devices such as the single electron transistor, where the stray capacitance $C_s$ is typically less than 1 pF. In this case matching is achieved at large resistances without having to use a large inductance $L$, yielding high resonance frequency and hence a large bandwidth. As an example, we plot simulated $S11$ traces ($S11$ is the ratio of the reflected power to the input power in dB) using $L=300$ nH and two different stray capacitances in Figure \ref{fig:basicRLC}. For low capacitance, $C_s=0.3$ pF, the resonance frequency is $f_c\approx$ 530 MHz. At the resonance frequency $S11$ is very sensitive to the resistance $R$; changing $R$ from 20 k$\Omega$ to 100 k$\Omega$ changes $S11$ by over 70 dB.

If the capacitance to ground is larger, $C_s=30$ pF, as would be the case in a large area device e.g., a gated 2D system, the resonance frequency drops to $\sim$46 MHz. Not only does this reduce the bandwidth, it also affects the matching so that although there is a 70 dB change in $S11$ at resonance when $R$ changes from 200 $\Omega$ to 1000 $\Omega$, there is less than 3 dB change when $R$ is increased from 1 k$\Omega$ to 100 k$\Omega$. The simulation with $C_s=30$ pF also gives a naive prediction for the sensitivity of the rf reflectometry technique using a standard Hall bar with a top-gate a few hundred nm above the 2D system. %, such that $C_g$=30 pF.
In such a device the resistance varies from a few k$\Omega$ up to as high as one can measure, so one would not expect rf reflectometry to be a sensitive technique for 2D systems in standard large area Hall bar devices. The matching resistance can be increased with a larger inductance $L$, but to have $R_m=20$ k$\Omega$ with $C_s=30$ pF would require $L=12\mu$H. Surface mounted inductors with such large inductance have a self resonance at few tens of megahertz hindering the measurement. In addition, if one needs to have a large bandwidth in the overcoupled regime, $L$ should be kept as small as possible because the bandwidth scales as $1/L$.

Contrary to these expectations, we show that rf reflectometry can be used for fast resistance measurement in gated 2D devices that are not specifically designed for high frequency measurements. Furthermore we show that for these samples we do not need to incorporate on-chip waveguides, or mount them in special rf packages.

\section{\label{exp} Experimental details}

\begin{figure}[ht]
\includegraphics[width=0.56\linewidth]{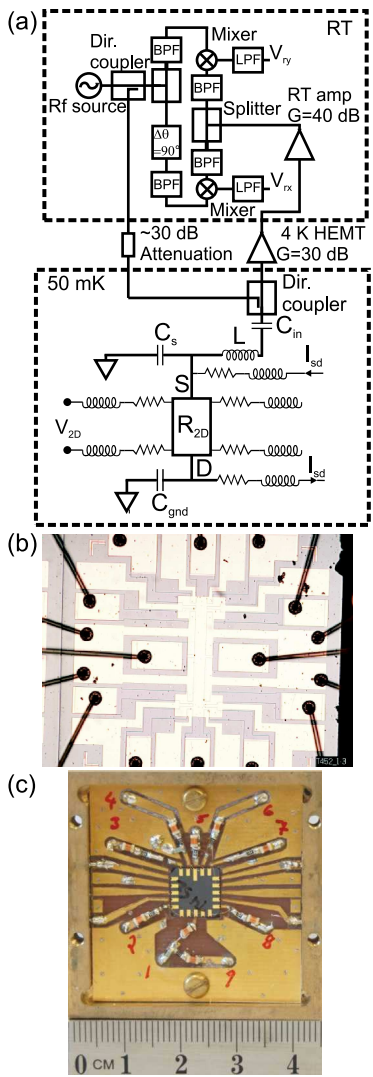}% Here is how to import EPS art
\caption{\label{fig:sampleholder} (a) Schematic of the rf setup. The carrier wave to the sample is fed through a directional coupler, 30 dB of attenuation at 4 K and another directional coupler close to the sample holder. The reflected signal is amplified in two stages and split in two for demodulation in two mixers which have a 90$^\circ$ phase difference in the reference signals. (b) Micrograph of the measured Hall bar device. The source connection is at the top of the picture and the drain at the bottom. The white area on top of the Hall bar is the metal gate used to change the carrier density and hence the resistance $R_{2D}$. (c) Photograph of the sample holder with a LCC-20 chip carrier mounted at the middle of the PCB. The rf comes in through connection 1 via a surface mounted input capacitor on the PCB. The other connections, 2-9, have $RL$ chokes enabling simultaneous rf and low frequency a.c. lock-in measurements. }
\end{figure}

To test the idea of using reflectometry to measure the resistance of a large area device, we embedded a 2DHS into an $LC$ circuit mounted on the cold finger of a dilution refrigerator. The sample, shown in Figure \ref{fig:sampleholder}(b), is a high mobility 2D hole system formed in a 30 nm GaAs quantum well buried 270 nm beneath the surface of an AlGaAs/GaAs heterostructure etched to form a Hall bar geometry. A metal gate is deposited on top of the Hall bar and the gated part has dimensions of $80\:\mu m\times800\:\mu m$. The 2DHS has a density $p=5\times10^{10}$ cm$^{-2}$ and mobility $\mu=5\times10^5$ cm$^2$V$^{-1}$s$^{-1}$. Full details of the heterostructure can be found in \cite{arh01}.

The chip is mounted in a LCC-20 chip carrier (Spectrum Semiconductor). Figure \ref{fig:sampleholder}(c) shows a photograph of the PCB we have used with the chip carrier mounted face down in the center. Electrical connection between the PCB and LCC-20 is via 20 $\mu$m gold wires attached using an ultrasonic Ball bonder (Kulicke and Soffa). The impedance matching network, bias tee, and rf chokes are made using standard surface mounted components on the PCB. The schematic in Figure \ref{fig:sampleholder}(a) shows the connections used to allow four-terminal low-frequency a.c. measurements to be conducted simultaneously with the rf measurements. The key to this capability is a set of series $RL$ chokes connected to the source (S), drain (D) and four Hall probes of the device. These chokes are mounted on the PCB close to the device and have resistors $R_{ch}=10$ k$\Omega$ for the source, drain and gate contacts (not shown in Figure \ref{fig:sampleholder}(a)) and $R_{ch}=100$ k$\Omega$ for the Hall probe lines. The choke inductors are either $L_{ch}=220$ nH or $L_{ch}=$390 nH. The resistance of the 2D hole system, $R_{2D}=V_{2D}/I_{sd}$, was measured with a standard a.c. lock-in technique and four terminal (4T) setup using an excitation voltage of $V_{exc}=50$ $\mu$V and frequency of 13 Hz.  

The rf setup, shown in Figure \ref{fig:sampleholder}(a), is similar to that presented in \cite{bue04}, except there is only one tank circuit and one rf source. In the measurements the carrier frequency was 328.1 MHz and the rf power incident to the tank circuit was about -85 dBm. The rf output is fed into a directional coupler which is used as an attenuator ($\sim$-16 dB) for the carrier wave launched into the fridge through a coaxial cable. The remainder of the rf wave is used as a reference signal for the mixers, and it is split in two and band pass filtered before it is coupled to the local oscillator (LO) input of the mixers. The reference signal for the other mixer is phase shifted 90$^\circ$ to enable quadrature measurements of the reflected power.

The carrier wave directed to the refrigerator is fed through another directional coupler at low temperature and is coupled to the input of the tank circuit via a capacitor $C_{in}=470$ pF. The inductor $L=100$ nH for the impedance transformer was chosen such that the resonance frequency lies within the bandwidth of the 4 K HEMT amplifier (between 300 MHz and 400 MHz). The rf ground is provided through another capacitor $C_{gnd}=470$ pF at the drain end of the Hall bar. Both $C_{in}$ and $C_{gnd}$ capacitors are mounted on the PCB.

The reflected carrier wave is amplified in two stages. The first stage, a HEMT amplifier at 4 K, has gain $G\approx 30$ dB and the second stage at room temperature has $G=40$ dB. After the room temperature amplifier the carrier wave is split and band pass filtered before entering the mixers for demodulation. The intermediate frequency outputs of the mixers were low pass filtered and measured either with a fast 4 channel oscilloscope or the auxiliary inputs of an SR830 lock-in amplifier. From the demodulated components $V_{rx}$ and $V_{ry}$ one can calculate the total amplitude of the reflected carrier $V_r=\sqrt{V_{rx}^2+V_{ry}^2}$.

\section{\label{sec:char}Tank circuit characterisation and modelling}

\begin{figure}[h]
\includegraphics[width=0.9\linewidth]{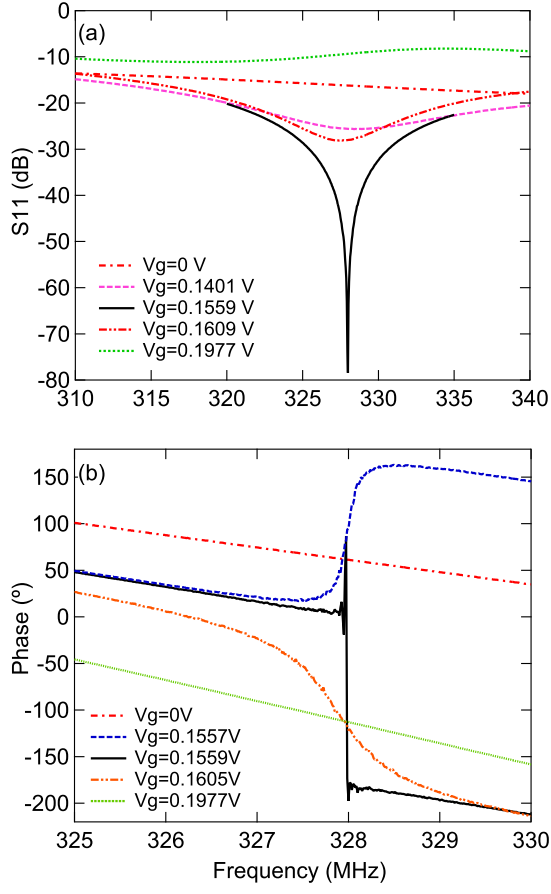}% Here is how to import EPS art
\caption{\label{fig:netw} (a) Measured rf reflectance $S11$ and (b) the phase difference between input and reflected signals for five gate voltages $V_g$.}
\end{figure}

We begin with a basic characterisation of the rf setup. Figure \ref{fig:netw}(a) shows the rf reflectance $S11$ measured at the output of the room temperature amplifier. At gate voltage $V_g=0$ V the resonance is very shallow. As $V_g$ is increased, the resistance $R_{2D}$ of the gated 2DHS increases, and the resonance becomes narrower and deeper. A sharp resonance appears at $V_g\approx$ 0.1559 V. At this gate voltage $R_{2D}\approx55$ k$\Omega$, which is equal to the matching resistance $R_m$ for this setup. The resonance frequency at the matching point is $f_c=328$ MHz. The measured $R_m$ and $f_c$ are much larger than predicted from the simple $RLC$ circuit (Fig. \ref{fig:basicRLC} inset) with nominal component values $L=100$ nH and $C_s=30$ pF, which would yield an expected resonance at 90 MHz with a matching resistance of 70 $\Omega$. Further increasing $V_g$, and hence $R_{2D}$, makes the resonance shallow and broad. The bandwidth of the resonant circuit can be roughly estimated from the -3 dB width of the resonance peak and exceeds 10 MHz in the current setup. We also note that there is a slight shift of the resonance towards lower frequencies as the device resistance is increased.

In the simple $RLC$ case, if we use the observed resonance frequency at the matching point, $f_c\approx 328$ MHz, and $L=100$ nH, we obtain a stray capacitance $C_s\approx 2.4$ pF, which is over ten times lower than we estimate for the geometric capacitance to the top gate alone. The expected matching resistance would in this case be $R_m=L/C_sZ_0=850$ $\Omega$, which is much lower than the measured 55 k$\Omega$.  

The phase difference between the input and the reflected carrier wave is plotted in Fig. \ref{fig:netw}(b). There is a 180$^\circ$ phase shift as we go through the resonance, and the shift becomes extremely sharp at the matching condition ($V_g\approx 0.1559$ V).  

\begin{figure}[h]
\includegraphics[width=0.98\linewidth]{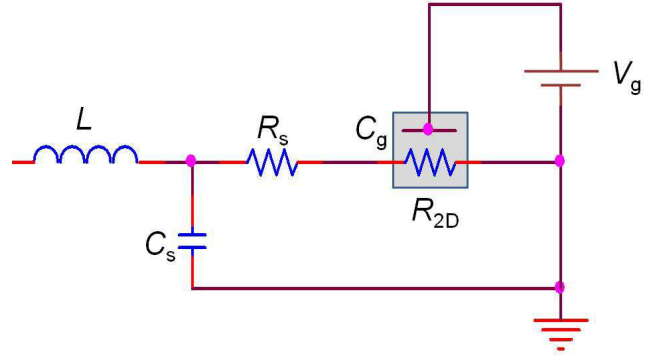}% Here is how to import EPS art
\caption{\label{fig:schema} Schematic of the lumped element model of the device and the tank circuit. $L$ is the inductance of the surface mounted component on the PCB, $C_s$ is the stray capacitance to ground, $R_s$ is the combined series resistance of the ohmics and the non-gated 2DHS, $C_g$ is the capacitance between the gate and the 2DHS, and $R_{2D}$ is the resistance of the gated 2DHS. A positive gate voltage $V_g$ can be applied to reduce the hole density and hence increase the resistance $R_{2D}$. }
\end{figure}

To understand why the gated 2D hole system does not behave as a simple resistor, as in Fig. \ref{fig:basicRLC} inset, we introduce the simple model shown in Fig. \ref{fig:schema}. The wavelength of the carrier at frequencies used here is sufficiently large to make a simple lumped element modelling accurate enough to give a qualitative understanding of the circuit response. 
The model consists of discrete components: The gated part of 2DHS is described by a resistive transmission line which is characterised by the resistance $R_{2D}$ and the gate capacitance $C_g$. The characteristic impedance of a generic transmission line is given by:
\begin{equation}\label{eq:trimpedance}
Z_{tr}=\sqrt{\frac{R_{tr}+j\omega L_{tr}}{G_{tr}+j\omega C_{tr}}},
\end{equation}
where $R_{tr}$ and $L_{tr}$ are the resistance and inductance per unit length and $C_{tr}$ and $G_{tr}$ are the shunt capacitance and conductance per unit length \cite{pozar}. In our device $j\omega L_{tr}$ and $G_{tr}$ are at least three orders of magnitude smaller than $R_{tr}$ and $j\omega C_{tr}$, respectively, and since the impedance $Z_{tr}$ depends on the ratio of the resistance and capacitance we can substitute $R_{tr}$ and $C_{tr}$ with the resistance of the gated 2DHS $R_{2D}$ and the gate capacitance $C_g$ (or with resistance per square $\rho_{xx}$ and capacitance per square $C_{sq}$). The characteristic impedance can then be written as: 
\begin{equation}\label{eq:trimpedance2}
Z_{tr}\approx \sqrt{R_{2D}/j\omega C_{g}}=\sqrt{\rho_{xx}/j\omega C_{sq}}.
\end{equation}
%, as depicted in figure \ref{fig:schema}. 

\begin{figure}[h]
\includegraphics[width=0.92\linewidth]{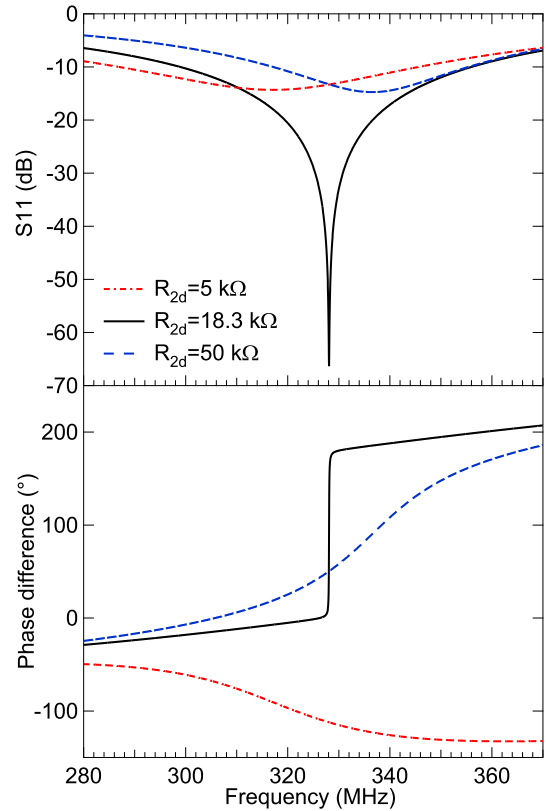}% Here is how to import EPS art 
\caption{\label{fig:simu} Calculated (a) $S11$ and (b) phase difference between injected and reflected signals.}
\end{figure}

The gated 2DHS is in series with the resistance $R_s$, which takes into account the ohmic contacts as well as non-gated regions of the 2DHS. Using Eqn. \ref{eq:trimpedance2}, we can obtain the combined impedance of the tank circuit and the sample:
\begin{equation}\label{eq:impedance}
Z_{t}=j\omega L+\frac{1}{\left(R_s+(1-j)\sqrt{R_{2D}/2\omega C_g}\right)^{-1}+j\omega C_s}.
\end{equation}
Using Eqn. \ref{eq:impedance} we can calculate the reflection coefficient $\Gamma$ and $S11=20\log{\vert\Gamma\vert}$. In Fig. \ref{fig:simu}(a) we plot three calculated $S11$ traces , where we have used a nominal component value $L=100$ nH, an estimated gate capacitance $C_g=30$ pF and series resistance $R_s=300$ $\Omega$. The stray capacitance was set to $C_s=1.92$ pF so that a resonance frequency of $f_c=328$ MHz is obtained at matching condition (as in the experiment - see Fig. \ref{fig:netw}). In this case, the matching occurs at $R_{2D}=18.3$ k$\Omega$. This is somewhat less than the measured matching resistance of 55 k$\Omega$ but not unreasonable given the simplicity of the model. For example, there are connections from the 2DHS to the measurement setup which are used for four terminal resistance measurements, and some leakage of the rf signal to ground can occur through these connections. This is not taken into account in the model. However, the shape of the calculated trace with $R_{2D}=18.3$ k$\Omega$ is very similar to that measured at the matching point (trace at $V_g=0.1559$ V in Fig. \ref{fig:netw}(a)). As $R_{2D}$ is changed from the matching value, the resonance gets wider and shallower. 

The calculated phase of the reflected rf signal is plotted in Fig. \ref{fig:simu}(b). The phase looks similar to the data shown in Fig. \ref{fig:netw}(b), although it changes over a wider frequency span than in the measured data: the 180$^\circ$ phase shift at the matching condition is very sharp, gets smoother away from the matching condition, and inverts as the system goes from the undercoupled $R_{2D}<R_m$ to the overcoupled $R_{2D}>R_m$ regime. 

Although the model is very simple, it is possible to grasp why the reflectometry method works for large gated 2D systems. The key is that the large gate capacitance reduces the total impedance of the Hall bar, making the method sensitive even at large 2DHS resistances where the simple RLC model predicts there should be almost no sensitivity. In addition, a considerable series resistance between the gated part of the device and the tank circuit makes the resonance frequency higher than otherwise would be the case. 

It should be noted that the setup is quite sensitive to the way the 2DHS is connected to the measurement circuit. For example, if we increase the number of ohmic contacts bonded from the source to the RF-in connector, we find that the measured resonance frequency is lower and that matching occurs at lower $R_{2D}$. The change in bonding also affects how the resonance frequency changes with $R_{2D}$. When we have three ohmic contacts bonded to the RF-in connector there is a slight downward shift in the resonance frequency with increasing $R_{2D}$, as seen in Fig. \ref{fig:netw}. However, when a single ohmic contact is bonded to the RF-in connector, the resonance frequency shifts upward with increasing $R_{2D}$, as it does in the calculated traces.

\section{\label{sec:resmeas}Reflectometry as a probe of 2DHS resistance}
We have demonstrated by a simple model and by measuring $S11$ that the reflection coefficient $\Gamma$ can be modulated by varying the gate voltage and hence the resistance of the 2DHS. We now measure the amplitude of the reflected rf carrier wave at 328.1 MHz, where sensitivity to the changes in $R_{2D}$ is large. By mixing the amplified reflected signal with the reference signal (see Fig. \ref{fig:sampleholder}(a)) we obtain the dc voltages $V_{rx}$ and $V_{ry}$ (where $V_{rx}$ is 90$^\circ$ phase shifted with respect to $V_{ry}$), and $V_r=\sqrt{V_{rx}^2+V_{ry}^2}$, which is proportional to the total reflected power. We compare the reflected rf signal to the 2DHS resistance $R_{2D}$ and show that the rf reflectometry technique can be used as a sensitive high bandwidth, low noise probe of the 2DHS resistance.

\begin{figure}[h]
\includegraphics[width=0.99\linewidth]{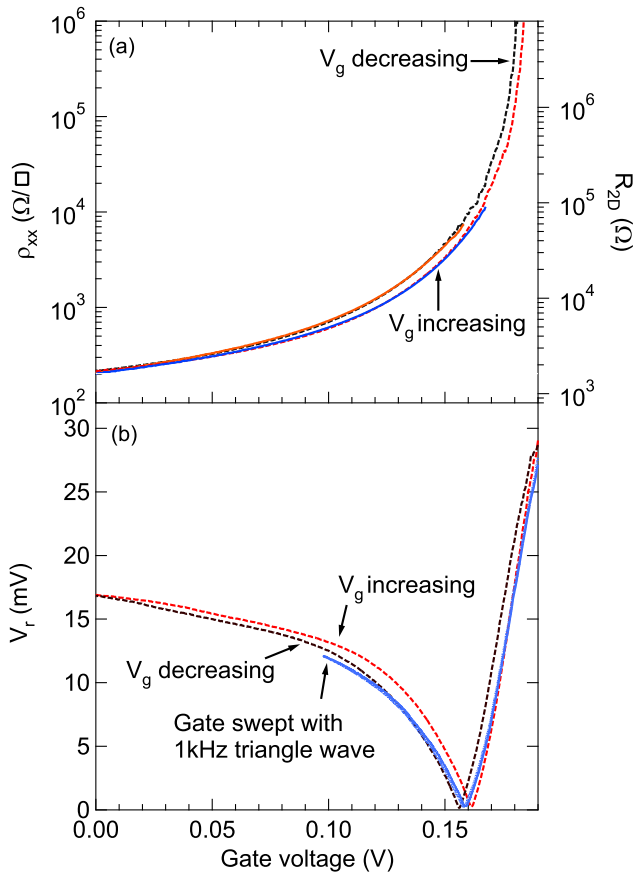}% Here is how to import EPS art
\caption{\label{fig:RvsVg} (a) The measured four terminal resistivity (solid lines) and the two terminal resistance where the series resistance has been taken into account (dotted lines). The gate voltage was changed in discrete steps of 0.5 mV when increased and in steps of 0.4 mV when decreased. (b) The total mixer output, $V_r$, versus the gate voltage. The gate voltage was changed both in discrete steps (dashed lines), and smoothly between $V_g=0.1$ V and $V_g=0.19$ V using a 1 kHz triangle wave (solid lines).}
\end{figure}

Figure \ref{fig:RvsVg}(a) shows a four terminal measurement of $\rho_{xx}$ versus the gate voltage. The gate voltage was changed in discrete steps of 0.5 mV when it was increased and in steps of 0.4 mV when decreased. The resistance increases rapidly as $V_g$ is increased and the hole density decreases. We cannot measure $\rho_{xx}$ beyond $\sim$ 10 k$\Omega/\square$, as the voltage probes pinch off at $V_g\approx 0.167$ V and $V_g\approx 0.158$ V when the gate voltage is changed towards or away from pinch off, respectively. The dashed lines in Fig. \ref{fig:RvsVg}(a) are the two terminal data after the large series resistance in the rf chokes (230 k$\Omega$ in total at 50 mK) has been subtracted. We also see some hysteresis in the $\rho_{xx}$ versus $V_g$ data depending on the sweep direction. This is a known problem in this type of hole device \cite{dan98}.

In comparison, Fig. \ref{fig:RvsVg}(b) shows the rf response as the gate voltage is changed. The total mixer output $V_r$ decreases with increasing $V_g$ until the matching point is reached at $V_g\approx 0.161$ V (0.156 V) when $V_g$ is stepped towards (away from) pinch off. Above the matching condition, $V_r$ increases rapidly with $V_g$. It is evident that we see a similar sweep direction dependent hysteresis in $V_r$ as we see in $\rho_{xx}$. The rf measurement is sensitive up to a resistance of several M$\Omega$, as can be seen by comparing the plots in Fig. \ref{fig:RvsVg}(a) and (b).

To see whether we can follow the resistance while changing the gate voltage at a faster rate, we applied a gate voltage sweep using a signal generator. The solid line in Fig. \ref{fig:RvsVg}(b) between $V_g=0.1$ V and $V_g=0.19$ V is a measurement where the gate was swept with a 1 kHz triangle wave and the mixer outputs and the gate voltage were measured using a fast 4 channel oscilloscope. Now the matching occurs at $V_g\approx 0.159$ V (0.158 V), when the gate voltage is swept towards (away from) the pinch off (i.e. we see a smaller hysteresis when we apply a smooth voltage sweep on the gate than in the measurement where the gate voltage is changed in discrete steps). The small differences in the stepped and fast gate sweeps are most likely due to a slow relaxation that we see in $\rho_{xx}$ in this type of hole device \cite{lt08}.

\begin{figure}[h]
\includegraphics[width=0.9\linewidth]{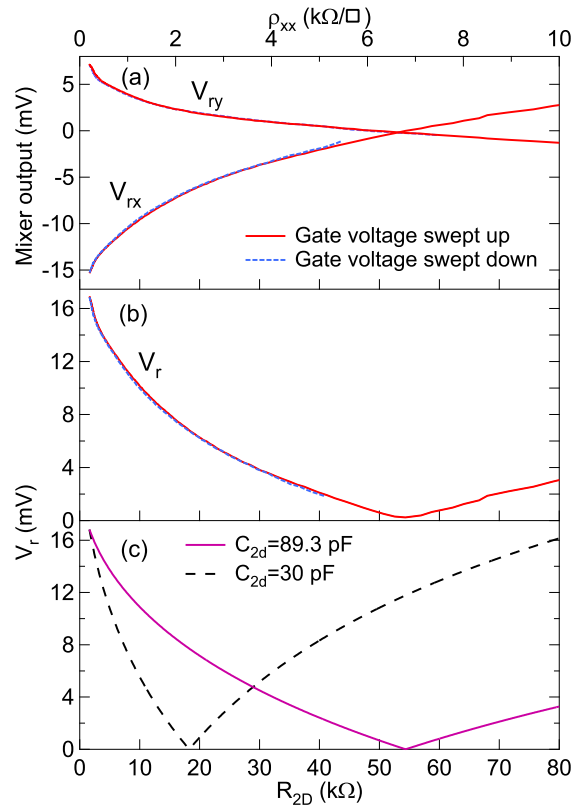}% Here is how to import EPS art
\caption{\label{fig:R} (a) The demodulated components from the mixer output $V_{rx}$ and $V_{ry}$, (b) the total amplitude of the mixer output $V_r$ and (c) the calculated transfer curves using $f=328.1$ MHz, $R_s=300$ $\Omega$, $L=100$ nH and $C_s=1.92$ pF. We have used two gate capacitances, the estimated gate capacitance $C_g=30$ pF, and $C_g=89.3$ pF to obtain the same matching resistance $R_{2D}=55$ k$\Omega$ as in the measurements.}
\end{figure}

The vital question is whether we can reliably map the 2DHS resistance to the amplitude $V_r$ of the reflected carrier wave. The transfer curve for our measurement setup is shown in Fig. \ref{fig:R}, where we have plotted the mixer outputs $V_{rx}$ and $V_{ry}$ in (a) and the total amplitude $V_r$ in (b) versus the four terminal resistance. The mixer outputs and four terminal resistance were measured simultaneously while the gate voltage was stepped slowly. The components $V_{rx}$ and $V_{ry}$ approach zero as the 2DHS resistance approaches the matching value, and cross each other just after the minimum in $V_r$. The amplitude $V_r$ decreases by two orders of magnitude as $\rho_{xx}$ is changed from 204 $\Omega/\square$ at $V_g=0$ V to 6.9 k$\Omega/\square$ at $V_g\sim 0.16$ V where the minimum is achieved ($R_{2D}$ changes from 1630 $\Omega$ to 55 k$\Omega$). After this point $V_r$ increases again, while $V_{ry}$ and $V_{rx}$ monotonically decrease and increase, respectively. Most significantly the transfer function does not have any hysteresis, making it possible to map the total mixer output $V_r$ to the resistance of the gated 2DHS. This provides a calibration that can be used to study the physics of 2D systems, for example, relaxation processes and shot noise. In practice, the monotonic behaviour in $V_{ry}$ and $V_{rx}$ versus $\rho_{xx}$ makes them more useful for determining $\rho_{xx}$ than $V_r$.

To compare the model and the measured results, we show the calculated transfer curves in Fig. \ref{fig:R}(c). We have scaled the calculated $\Gamma$ so that we get the same value for the calculated and measured $V_r$ at $R_{2D}=1630$ $\Omega$. The calculated curve for the estimated gate capacitance $C_g=30$ pF has qualitatively the same features as the measured one, but the matching resistance, 18.3 k$\Omega$, is smaller. When we increase the gate capacitance to $C_g=89.3$ pF, we get the matching at the same $R_{2D}$ as in the measurements (55 k$\Omega$), and the calculated transfer curve imitates the measured curve surprisingly well. The simplicity of the model means that we cannot draw any conclusions regarding the factor of three increase in the capacitance required to match the experimental data. Nevertheless the measurements and modelling give confidence that rf reflectometry can be used as a sensitive and fast probe of the resistance of a gated 2DHS. Furthermore, since the impedance of the gated 2DHS $Z_{tr}$ depends on the ratio of 2DHS resistance and gate capacitance, it is possible to tune $Z_{tr}$ at a given $R_{2D}$ by changing the device geometry, for example one can increase the matching resistance by fabricating wider Hall bars.  

\section{\label{sec:bfield}Mapping the Landau level evolution}
Applying a sufficiently large perpendicular magnetic field to a high quality 2DHS brings it into the quantum Hall regime. In this regime, as the density of the holes $p$ is changed the Hall resistance $R_H$ becomes quantised, and the longitudinal resistance $\rho_{xx}$ goes to zero every time a Landau level is filled \cite{qhe}. Therefore, the two terminal resistance of the Hall bar shows a minimum, and rf reflectometry should be a sensitive method to map out the evolution of the Landau levels as a function of the magnetic field and gate voltage. In a plot of the magnetic field $B$ versus the hole density $p$, the position of the minimum in $\rho_{xx}$ is given by:
\begin{equation}\label{eq:qhe}
B=ph/\nu e,
\end{equation}
where $p$ is the hole density and $\nu$ is the filling factor \cite{qhe}.

To test whether we can map out the evolution of the Landau levels using the reflectometry method, we applied a triangle wave at 1 kHz sweeping the gate between 0 V and 0.096 V. The magnetic field was swept slowly at a rate of 0.04 T/min from 0 T to 2 T. The mixer outputs and the gate voltage were measured using an oscilloscope. 

\begin{figure}[h]
\includegraphics[width=0.99\linewidth]{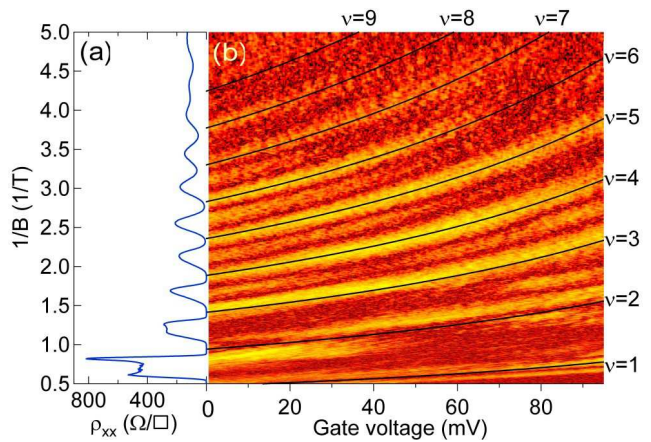}% Here is how to import EPS art
\caption{\label{fig:fastS11} (a) Resistivity $\rho_{xx}$ of the 2DHS at $V_g=0$ V versus the inverse magnetic field 1/$B$. (b) $V_{ry}$ measured while the magnetic field was varied slowly (0.04 T/min) and the gate voltage was swept between 0-0.96V with a 1 kHz triangle wave. We have plotted the logarithm of difference in the reflected signal $\log{\vert\Delta V_{ry}\vert}$ as a function of gate voltage and 1/$B$ to show the higher filling factor Landau levels more clearly. Black lines are the estimated Landau level evolution for $\nu=1$ - 9 using Eqn. \ref{eq:qhe}.}
\end{figure}

The two terminal nature of the rf reflectometry technique causes some complications in this type of measurement, since both the non-gated and gated sections of the Hall bar affect the rf reflectivity, and both have different, but strongly magnetic field dependent, resistance. However, by calculating the difference between the data at two gate voltages that are 2 mV apart, $\Delta V_{ry}(V_g)=V_{ry}(V_g-1mV)-V_{ry}(V_g+1mV)$, we can eliminate the modulation due to the non-gated region. To further improve the contrast we take the logarithm of the absolute value of the difference, $\log{\vert\Delta V_{ry}\vert}$, and plot it in Fig. \ref{fig:fastS11}(b) versus the inverse magnetic field to make the higher filling factors more visible.  

When we compare the $\rho_{xx}$ measured at $V_g=0$ V using standard low frequency lock-in techniques (Fig. \ref{fig:fastS11}(a)), to the color map, we see that the bright yellow features at $V_g=0$ V fit well to the minima in $\rho_{xx}$. As the gate voltage is increased the bright features move to lower magnetic fields, i.e. to larger 1/$B$, tracking the Landau level evolution. We have fitted the continuous black lines on top of the color map using Eqn. \ref{eq:qhe} assuming a linear dependence between the gate voltage and the hole density $p=(5.13-21.3V_g)\times 10^{14}m^{-2}$. The larger filling factors $\nu=3$ - 9 fit well to the features in the data. Two of the lowest filling factors do not fit as well, probably because the measurement at higher magnetic fields suffers more from the large resistance of the non-gated 2DHS. There are also fractional quantum Hall states starting to form close to $\nu=1$ and $\nu=2$, which further complicate the behaviour at low filling factor.

The slope ${\rm d}p/{\rm d}{V_g}=21.3\times 10^{14}V^{-1}m^{-2}$ can be used to estimate the effective distance between the gate and the 2DHS. If one assumes that the 2DHS and gate form a parallel plate capacitor, one obtains $d=\epsilon_0\epsilon_r/e{\rm d}p/{\rm d}V_g=330$ nm, where we have used the static dielectric constant $\epsilon_r=12.9$ for GaAs. The result is close to nominal growth specification of $d=285$ nm.

\section{\label{sec:conc}Conclusions}
We have shown that it is possible to embed gated large area 2D systems in an impedance transformer $LC$ circuit and use the amplitude modulation in the reflected carrier wave to measure temporal changes in the resistance of these systems. We have developed a simple model which gives a reasonable qualitative picture of why the method works better than the naive estimations would suggest. That is, the large 2D system can actually be thought of as a resistive transmission line with much smaller characteristic impedance than that of the 2D system alone.

This method can be used to speed up measurements since the gate voltage can be swept faster, as we have shown by mapping the Landau level evolution. It is also possible to measure resistance relaxation processes, such as glassy dynamics \cite{jar06, lt08}, at time scales many orders of magnitude shorter than is possible using conventional lock-in techniques. Moreover, studies of low frequency 1/$f$ resistance fluctuations \cite{let03, jar02} and shot noise in $\mu$m scale 2D systems, e.g. close to metal-insulator transition transition, could benefit from this method. 

\begin{acknowledgments}
The authors wish to acknowledge R. G. Clark for use of rf equipped dilution refrigerator and D. Barber for technical support. This work was supported by the Australian Research Council (Grant No. DP0558769).
\end{acknowledgments}

\bibliography{RfOn2D_cl}% Produces the bibliography via BibTeX.

\begin{thebibliography}{17}
\bibitem[{lt0()}]{lt08}
\bibinfo{note}{Results on slow resistance relaxation will be published
  elsewhere}.
\end{thebibliography}

\end{document}